\documentclass{emulateapj}
\usepackage{txfonts}
\usepackage{bm}
\usepackage{natbib}
\usepackage{graphicx}
\usepackage{psfrag}
\usepackage{pst-grad}

\newcommand{\eqb}{\begin{eqnarray}}
\newcommand{\eqe}{\end{eqnarray}}
\newcommand{\diff}{\textrm{d}}
\newcommand{\accel}{{a^*}}
\newcommand{\Ehat}{\hat{E}}
\newcommand{\Doppler}{{\cal D}}
\newcommand{\lambdaC}{\lambdabar_{\rm C}}
\newcommand{\rstar}{r_*}
\newcommand{\Pdot}{\dot{P}_{15}}
\newcommand{\rlight}{r_{\rm L}}
\newcommand{\omegalin}{\hat{\omega}_{\rm linac}}
\newcommand{\omegacurv}{\hat{\omega}_{\rm curv}}
\newcommand{\rhostar}{\rho_*}
\newcommand{\Bhat}{\hat{B}}
\newcommand{\evac}{e_{\rm vac}}
\newcommand{\espace}{e_{\rm space}}

\slugcomment{Submitted to ApJ}
\shorttitle{Linear acceleration emission}
\begin{document}

\title{Linear acceleration emission in pulsar magnetospheres}

\author{
Brian Reville
and John G. Kirk}
\affil{Max-Planck-Institut f\"ur Kernphysik, Postfach 10~39~80,
69029 Heidelberg, Germany}
\email{brian.reville@mpi-hd.mpg.de, john.kirk@mpi-hd.mpg.de}

\begin{abstract}
  Linear acceleration emission occurs when a charged
  particle is accelerated parallel to its velocity. We evaluate the
  spectral and angular distribution of this radiation for several
  special cases, including constant acceleration (hyperbolic
  motion) of finite duration. Based on these results, we find the
  following general properties of the emission from an electron in a
  linear accelerator that can be characterized by an electric field
  $E$ acting over a distance $L$: (i) the spectrum extends to a
  cut-off frequency $\hbar\omega_{\rm c}/mc^2\approx L \left(E/E_{\rm
      Schw}\right)^2/\lambdaC$, where $E_{\rm Schw}
  =1.3\times10^{18}\,\textrm{V\,m}^{-1}$ is the Schwinger critical
  field and $\lambdaC=\hbar/mc=3.86\times10^{-13}\,\textrm{m}$ is
  the Compton wavelength of the electron. (ii) the total energy emitted
  by a particle traversing the accelerator is $\frac{4}{3}\alpha_{\rm f}\hbar\omega_{\rm c}$,
  in accordance with the standard
  Larmor formula, where $\alpha_{\rm f}$ is the fine-structure constant. 
  (iii) the low frequency spectrum is flat for
  hyperbolic trajectories, but in general depends on the details of
  the accelerator. We also show that 
  linear acceleration emission complements curvature
  radiation in the strongly magnetized pair formation regions in
  pulsar magnetospheres.  It dominates when the length $L$ of the
  accelerator is less than the formation length $\rho/\gamma$ of
  curvature photons, where $\rho$ is the radius of curvature of the
  magnetic field lines, and $\gamma$ the Lorentz factor of the emitting
  particle. In standard static models of pair creating
  regions linear acceleration emission is negligible, but it is
  important in more realistic dynamical models in which the
  accelerating field fluctuates on a short length-scale.
\end{abstract}


\section{Introduction}

Linear acceleration emission is the radiation produced when a charged
particle is accelerated in a direction parallel to its velocity. The
resulting rectilinear trajectory is a special case. In astrophysics,
it is expected to occur in the inner parts of a pulsar magnetosphere,
where the strong magnetic field enforces one-dimensional motion on the
electrons, and there is a non-zero electric field aligned with the
magnetic field.  The calculation of linear acceleration emission is
complicated by the fact that relatively large sections of the
trajectory radiate coherently (in a sense that will be made more
precise in Section~\ref{equations}). In the celebrated case of \lq
hyperbolic\rq\ motion, in which the acceleration measured in the
instantaneous rest frame of the particle is constant, the entire
trajectory remains coherent and it is impossible to define a local
photon emission rate. Linear acceleration emission is a potentially
important component of the electromagnetic cascades that occur in
pulsars. These cascades are responsible for the
materialization of the electrons and positrons that radiate in pulsar
wind nebul\ae, and, ultimately, make their way to Earth as a
high-energy cosmic-ray component \citep[e.g.,][]{profumo09}.

Early work on this problem in the context of pulsars treated the {\em
  relativistic oscillator}, in which the electric field is a linear
function of position \citep{wagoner69}. Linear solutions were
considered by \citet{melrose78} and \citet{kuijpersvolwerk96}
who emphasized linear acceleration emission masers.
Recently, the focus has shifted to superluminal, large-amplitude electrostatic
waves. The particle trajectory and resulting emission in these waves was 
calculated using a
semi-classical formalism by \citet{rowe95} and has been
re-analyzed by \citet{melroseetal09} and \citet{melroseluo09} who
found and discussed some significant differences with Rowe's results.

In this paper we compute the emissivity for linear acceleration in
several special cases. The results enable us to estimate the emission
in the general case, and we use this estimate to discuss the
importance of the process in both static and dynamic models of the
pair producing regions in pulsar magnetospheres, which we call \lq\lq
gaps\rq\rq. Our method differs from that of \citet{melroseetal09} and
\cite{melroseluo09} in several respects. In particular, by avoiding
the use of the \lq Airy integral approximation\rq\ we resolve a puzzle
they encountered concerning the applicability of Larmor's formula to
linear acceleration. We also drop the restriction to superluminally
propagating disturbances, which enables us to treat static gaps, as
well as those containing spatially isolated electric field structures
such as double layers. Although the method and the details of the
emissivity we find are substantially different, the main conclusions
concerning the importance of linear acceleration emission in pulsars
are consistent with those of \citet{melroseluo09}: it is unimportant
for the parameters currently under discussion in static gap models of
pulsars, but could be the dominant radiation mechanism in dynamic gap
models.

The paper is organized as follows: in section~\ref{equations} we
briefly review the formalism of classical electrodynamics concerning
the radiation of test particles in prescribed vacuum fields,
emphasizing the role of the photon formation or coherence length. In
section~\ref{hyperbolic} we apply this formalism to the well-known
problem of hyperbolic orbits (motion with constant acceleration in the
particle's rest frame) and derive a useful approximate representation
of the angular and spectral distribution of radiated photons. In
section~\ref{special} we use the method of steepest descents to
compute an approximate emissivity for a simple wave form that can
describe a stationary or moving double layer.  We also present an
exact analytic solution for the spectrum emitted by a particle
undergoing simple harmonic motion.  In each case we demonstrate that
the results agree with simple estimates based on the spectrum found
for hyperbolic orbits.  Finally, we discuss in section~\ref{pulsars}
the role of linear acceleration and curvature emission in pulsar gap
models, comparing the relative importance of each process. Our
conclusions are presented in section \ref{conclusions}.

\section{Emissivity}
\label{equations}

Consider a particle that moves in vacuum. The spectral 
and angular distribution of the 
energy $U$ it 
radiates is \citep[e.g.,][]{schwinger98}
\eqb
\frac{\diff U}{\diff\omega\diff\Omega}&=&
\frac{\omega^2}{4\pi^2 c^3}
\left|\bm{n}\times\bm{j}(\omega\bm{n}/c,\omega)\right|^{2}
\label{emissivity}
\eqe
where $\bm{n}$ is a unit vector in the direction of propagation of the 
emitted wave, and $\bm{j}(\bm{k},\omega)$ is the Fourier
transform of the current:
\eqb
\bm{j}(\bm{k},\omega)&=&
\int\diff^3\bm{x}\int_{-\infty}^{\infty}\diff t\,\bm{j}(\bm{x},t)
\textrm{exp}\left(i\omega t-i\bm{k}\cdot\bm{x}\right)
\nonumber\\
&=&
q c\int_{-\infty}^{\infty}\diff t\,\bm{\beta}(t)
\textrm{exp}\left[i\phi(t)\right]
\enspace.
\eqe
Here, 
\eqb
\phi(t)&=&\omega t-\bm{k}\cdot\bm{x}(t)
\enspace,
\eqe
$\bm{x}(t)$ and $c\bm{\beta}(t)$ are the position and velocity of the
particle at time $t$. The quantity $\phi(t)$ can be identified as the
phase of the emitted wave at the particle's position, given that
$\phi=0$ at $t=0$. 
Thus,
\eqb
\frac{\diff U}{\diff\omega\diff\Omega}&=&
\frac{\omega^2q^2}{4\pi^2 c}
\int_{-\infty}^\infty
\diff t
\int_{-\infty}^\infty
\diff \tau
\left[\bm{n}\times\bm{\beta}(t)\right]\cdot
\left[\bm{n}\times\bm{\beta}(t+\tau)\right]
\nonumber\\
&&
\exp\left\lbrace -i\omega\tau+i
\omega\bm{n}\cdot\left[\bm{x}(t)-\bm{x}(t+\tau)\right]/c\right\rbrace
\label{correlation}
\eqe

For sufficiently large $\omega$, the exponential in (\ref{correlation})
is sharply peaked around $\tau=0$, which implies that the
$\tau$-integral can be written as a function of the particle velocity
and its derivative at time $t$, i.e., as the integral over the
trajectory of a locally defined emissivity. For lower frequencies,
however, this is not the case, and different parts of the trajectory
contribute coherently to the emission. At any time $t$, the section of
the trajectory that contributes coherently is that over which the
phases $\phi(t)$ and $\phi(t+\tau)$ do not differ by a large amount;
i.e., one can identify the time $\tau_{\rm coh}$ over which $\phi(t)$ changes
by $2\pi$ as a characteristic coherence time, or photon formation time
 \citep[e.g][]{akhiezer87}. 

 For a relativistic particle whose trajectory has a radius of
 curvature $R$, the coherence time is approximately
 $\textrm{Max}(\gamma^2/\omega,R/\gamma c)$, and a
 local emissivity can be defined, provided that the accelerating
 fields can be considered constant on this timescale. This is the
 normally encountered case of synchrotron radiation. However, for both
 \lq jitter\rq\ radiation
 \citep[e.g.,][]{fleishmantoptygin,medvedev,kirkreville10} and linear
 acceleration emission, the coherence time is not short compared
 to the timescale on which the accelerating fields vary.

For a particle that experiences acceleration only 
during a finite time interval,
$t_1<t<t_2$,
the Fourier transform of the current 
can 
be written in terms of the trajectory within this interval:
\eqb
\bm{j}(\bm{k},\omega)&=&iqc\int_{t_1}^{t_2}\diff t\,
\textrm{e}^{i\phi(t)}\frac{\diff}{\diff t}\left(
\frac{\bm{\beta}(t)}{\diff\phi/\diff t}\right)
\label{schwinger1}
\enspace,
\eqe
as emphasized by \citet{schwinger98}.
Specializing to the case of linear acceleration emission in vacuum,
$\bm{k}=\bm{n}\omega/c$ and $\bm{n}\cdot\bm{\beta}=\mu\beta$ and the
cosine $\mu$ of the angle between $\bm{n}$ and
$\hat{\bm{x}}=\bm{\beta}/\beta$ is a constant.
Therefore
\eqb
\bm{j}(\bm{k},\omega)&=&\frac{i qc}{\omega}\hat{\bm{x}}
\int_{t_1}^{t_2}\diff t\,
\textrm{e}^{i\omega\left[t-\mu x(t)/c\right]}
\frac{\diff}{\diff t}\left[
\frac{\beta(t)}{1-\mu\beta(t)}\right]
\nonumber\\
&=&\frac{i qc}{\mu\omega}\hat{\bm{x}}
\left\lbrace
\left[\frac{\textrm{e}^{i\omega\left[t-\mu x(t)/c\right]}}{1-\mu\beta(t)}
\right]_{t_1}^{t_2}\right.
\nonumber\\
&&\left.
-i\omega\int_{t_1}^{t_2}\diff t\,
\textrm{e}^{i\omega\left[t-\mu x(t)/c\right]}\right\rbrace
\label{schwinger2}
\enspace,
\eqe
provided $\mu\neq0$.
The expressions (\ref{schwinger1}) and (\ref{schwinger2}) 
are the starting points of 
our analysis of the emissivity. It is worth noting that when (\ref{schwinger1})
is inserted into (\ref{emissivity}) the emissivity can immediately
be integrated over frequency and angle to give
\eqb
U&=&\frac{2 e^2}{3 c}\int_{t_1}^{t_2}\diff t\,\gamma^4\dot{\bm{\beta}}^2+
\gamma^6\left(\bm{\beta}\cdot\dot{\bm{\beta}}\right)^2
\nonumber\\
&=&
\int_{t_1}^{t_2}\diff t\,\frac{2 e^2}{3 m^2c^3}\left|
\frac{\diff p^\mu}{\diff\tau}\right|^2
\label{larmor}
\eqe
where $\dot{\bm{\beta}}=\diff\bm{\beta}/\diff t$,
$p^\mu=mc\left(\gamma,\gamma\bm{\beta}\right)$ is the particle momentum 
$4$-vector and $\tau$ the proper time. The integrand on the right-hand side of 
(\ref{larmor}) is just the relativistic form of Larmor's formula for the 
instantaneous power radiated. Thus, when integrated over a trajectory 
on which the acceleration is non-vanishing only within a finite time interval, 
this formula correctly gives the 
total radiated power. It is, nevertheless, incorrect to regard it
as a truly instantaneous power, for example by associating a particular
section of the trajectory with a part of the radiated power
\citep[e.g.,][Chap.~37]{schwinger98}.

\section{Hyperbolic motion}
\label{hyperbolic}

The trajectory of a 
particle moving with constant acceleration (as measured in its instantaneous
rest frame) is called \lq\lq hyperbolic\rq\rq\ because of its
shape in the $(x,t)$-plane (the acceleration and velocity 
are assumed parallel). 
In classical
electrodynamics in flat space, a hyperbolic trajectory results 
when a particle moves in a constant
electric field which is parallel to its velocity, and also to the magnetic 
field, if any is present. The sometimes controversial 
literature on the problem of the radiation emitted by a particle 
on such a trajectory,
goes back over a century
--- see, for example, \citet{ginzburg70}. In the last few decades,
interest has arisen in the associated quantum effects, the {\em Unruh Effect} 
and {\em Unruh Radiation} 
\citep[e.g.,][]{unruh76,bellleinaas87,crispinoetal08,thirolfetal09}.
At least in the classical limit, the conceptual problems 
associated with hyperbolic motion disappear if 
the particle experiences acceleration during only a finite time-interval,
outside of which it moves with constant velocity. We will assume this 
to be the case in the application to pulsars. 

Consider a particle of charge $q$ and mass $m$ 
moving along the $x$-axis in an 
electric field $\bm{E}(x)=\hat{\bm{x}}E(x)$ which is constant 
and of magnitude $E$ between
$x=0$ and $x=L$ and vanishes elsewhere. Assuming the velocity is also
directed along $\hat{\bm{x}}$, the trajectory is
\eqb
x&=&\accel\left[\cosh\left(c\tau/\accel\right)-1\right], \quad ct\,=\,\accel
\sinh\left(c\tau/\accel\right)
\label{hyperbola}
\eqe
for $0<\tau<\tau_L=
\left(\accel/c\right)\textrm{cosh}^{-1}\left[\left(L+\accel\right)/\accel\right]$.
The length $\accel$ is the distance over which the electrostatic potential
changes by $mc^2/|q|$: 
\eqb
\accel&=&\frac{mc^2}{\left|q\right|E}
\nonumber\\
&=&\lambdaC/\Ehat
\enspace,
\eqe
where $\lambdaC=\hbar/mc$ is the Compton wavelength and 
$\Ehat=|q|\hbar E/m^2c^3$ is the electric field in units of the critical
field --- for an electron or positron this is the Schwinger field $E_{\rm Schw}=1.3\times10^{18}
\,\textrm{V\,m}^{-1}$. The trajectory (\ref{hyperbola}) is an exact solution
of the classical equations of motion, including the Lorentz-Abraham-Dirac form
of the radiation reaction force (which vanishes identically for 
$0<\tau<\tau_L$).
 
In a gap near a pulsar surface, the magnetic field is sufficiently
strong that electrons and positrons rapidly decay into the Landau ground state.
Then, if the radius of curvature of the field lines is large 
(see section~\ref{pulsars}) the particle motion is approximately
one-dimensional. The
component along $\bm{B}$
of the electric field induced by rotation of the star accelerates particles
in a trajectory that can be approximated by (\ref{hyperbola}), whereas the 
perpendicular component produces an $\bm{E}\times\bm{B}$-drift that can
be transformed away. Static models of gaps assume the (parallel) field 
vanishes below the surface ($x=0$) and above 
a \lq\lq pair-production front\rq\rq\ at height $L$, 
where photons produced in the gap create
a sufficient number of charges to screen the field. Under pulsar conditions,
one generally expects $\Ehat\ll 1$ and $L/\accel\gg1$ (see Sect.~\ref{pulsars}). 

A particle 
that starts at rest at the surface and is accelerated upwards
achieves a Lorentz factor $\gamma=\left(x+\accel\right)/\accel$ at height $x$.
The energy radiated in passing from the surface to the pair production front
can be computed
from (\ref{schwinger1}) or (\ref{schwinger2}). 
For emission at $\bm{n}\cdot\hat{\bm{x}}=\mu$, a simple 
approximation can be found by first transforming into the 
particle rest frame at proper time 
$\tau_0=\left(\accel/c\right)\textrm{tanh}^{-1}\left(\mu\right)$,
when it has reached the position 
$x_0=\accel\left[\cosh\left(c\tau_0/\accel\right)-1\right]$ and a Lorentz factor
$\gamma_0=\left(x_0+\accel\right)/\accel$.
Denoting coordinates in this frame by $(x',t')$, the trajectory is
\eqb
x'&=&\accel\left\lbrace
\cosh\left[c\left(\tau-\tau_0\right)/\accel\right]-1\right\rbrace+x_0\enspace,\\
ct'&=&\accel
\sinh\left[c\left(\tau-\tau_0\right)/\accel\right]
\label{hyperbolaprime}
\eqe
For photons of frequency $\omega'$
emitted at $\mu'=0$, i.e., normal to $\hat{\bm{x}}$
in this frame, one finds
\citep{schwinger98}
\eqb
\label{SchwQ2}
\left.\frac{\diff U'}{\diff\omega'\diff\Omega'}\right|_{\mu' = 0}
&\approx&\frac{q^2}{\pi^2c}
\left[\omega'\frac{\accel}{c}K_1\left(\omega'\frac{\accel}{c}\right)
\right]^2
\label{mcdonald}
\enspace,
\eqe
where $K_1$ is a modified Bessel function of the second kind.
This approximate solution involves extending the limits of the 
integration over the trajectory to $\tau=\pm\infty$, and is accurate
provided the endpoints in the new (primed) frame of reference
are far from the point at 
which the particle is at rest: $0\ll\tau_0\ll\tau_L$.
Returning to the frame in which the stellar surface is at rest,
(\ref{mcdonald}) describes the radiation emitted at 
$\mu = \beta=\sqrt{\gamma^2-1}/\gamma$. Therefore, 
exploiting the transformations
$\omega'=\Doppler\omega$, $\diff U'=\Doppler\diff U$,
$\diff\Omega=\Doppler^{-2}\diff\Omega'$ where $\Doppler=\gamma(1-\mu\beta)
=\sqrt{1-\mu^2}$ is the 
Doppler factor,
\eqb
\left.\frac{\diff U}{\diff\omega\diff\Omega}\right|_{\mu=\beta}&=&
\left.\Doppler^2
\frac{\diff U'}{\diff\omega'\diff\Omega'}\right|_{\mu'=0}
\enspace,
\eqe
whence
\eqb
\label{HyperAllmu}
\frac{\diff U}{\diff\omega\diff\Omega}
&=&\frac{q^2}{\pi^2c}
\left[\omega\frac{\accel}{c}K_1\left(\omega\sin\theta\frac{\accel}{c}\right)
\right]^2
\eqe
where $\theta=\cos^{-1}\mu$.
The result can be checked by integrating over all 
frequencies, using the identity
\eqb
\int_0^\infty x^2  K^2_\nu(x)\diff 
x&=&\frac{\pi^2(1-4\nu^2)}{32\cos\pi\nu}
\eqe
to find
\eqb
\label{SchwQ3}
\frac{\diff U}{\diff\Omega}
&=&\frac{3q^2}{32\accel}\frac{1}{\left(1-\mu^2\right)^{3/2}}
\enspace,
\eqe
in agreement with the expression given by \citet{schwinger98}.
For small argument, $K_1(x)\approx1/x$, so that the low frequency 
spectrum is flat:
\eqb
\label{Hyperlowfreq}
\frac{\diff U}{\diff\omega\diff\Omega}
&=&\frac{q^2}{\pi^2c\left(1-\mu^2\right)}
\enspace.
\eqe
Using the asymptotic form of $K_1$ for large argument, one finds
the high frequency spectrum:
\eqb
\label{Hyperhighfreq}
\frac{\diff U}{\diff\omega\diff\Omega}
&=&\frac{q^2 }{4\pi c}\frac{\omega}{\omega_{\rm c}}
{\rm e}^{-\omega/\omega_{\rm c}}
\enspace,
\eqe
which has a cut-off at the
frequency 
\eqb
\omega_{\rm c}&=&\frac{c}{2\accel\sqrt{1-\mu^2}}
\enspace.
\eqe

As expected, for relativistic particles,
most of the energy is radiated close to the forward 
direction. The method applies only to radiation inside  
a small forwardly directed cone of opening angle $\sqrt{1-\mu^2}\ll1$,
because otherwise $x_0$ is too close to the stellar surface. 
However, the region outside of this cone is not
expected to contain significant power. The approximation also fails within
a very small, forwardly directed cone, because the appropriate value of 
$\tau_0$ approaches or exceeds $\tau_L$. This is potentially 
a more serious shortcoming, since it prevents 
one computing the emissivity on the 
beaming cone of a particle when it reaches its maximum Lorentz factor
$\gamma_{\rm max}=\left(L+\accel\right)/\accel$ at the pair production front. 

\begin{figure}
\includegraphics[width=0.45\textwidth]{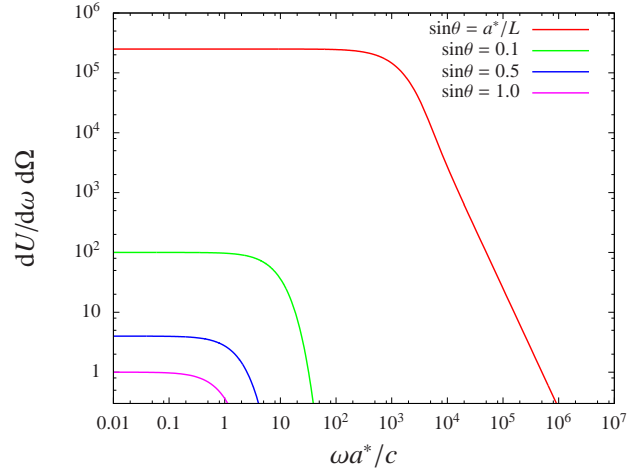}
\caption{
Energy spectrum, in units of $q^2/\pi^2c$, radiated at different angles using equation
(\ref{HyperAllmu}) for $\sin\theta=0.1,\ 0.5,\ 1.0$ and equation (\ref{Struve}) 
for $\sin\theta=\accel/L=10^{-3}$. Note that the high frequency
power law at angle $\theta=\accel/L$ is an artifact of the
discontinuity in the electric field (see text).
A color version of this figure is available in the online journal. \\
}
\label{spectrumK}
\end{figure}

In fact, an explicit expression
for the emission at $\mu=\sqrt{\gamma_{\rm max}^2-1}/\gamma_{\rm max}$ 
can be written down by 
evaluating the integral over the orbit between the limits $\tau=-\infty$ 
and  $\tau=\tau_L$: 
\eqb
\label{Struve}
\left.\frac{\diff U}{\diff\omega\diff\Omega}\right|_{\mu = 
\sqrt{L^2-\accel^2}/L}
&=&\frac{q^2\omega^2\accel^2}{4\pi^2c^3}\\
\left\lbrace K_1^2\left(\omega\frac{\accel^2}{Lc}\right)
\right.&+&\left.\frac{\pi^2}{4}
\left[I_1\left(\omega\frac{\accel^2}{Lc}\right)-
L_{1}\left(\omega\frac{\accel^2}{Lc}\right)
-\frac{2}{\pi}\right]^2\right\rbrace
\nonumber
\eqe
where $I_1$ is the modified Bessel function of the first kind and 
$L_{1}$ is
the modified Struve function \citep{AbramowitzStegun}. The spectrum at different 
angles, using this formula and equation (\ref{HyperAllmu}) is shown in figure 
\ref{spectrumK}.
At high frequencies
$\omega\gg Lc/\accel^2$ one finds from (\ref{Struve}) the asymptotic expression
\eqb
\left.\frac{\diff U}{\diff\omega\diff\Omega}\right|_{\mu = \sqrt{\gamma_{\rm max}^2-1}/\gamma_{\rm max}}
&=&\frac{q^2 L^2 c}{4\pi^2\omega^2\accel^4}\enspace,
\eqe

In contrast with the emission at larger angle, 
this spectrum does not exhibit an exponential cut-off, but
is a power law $\propto\omega^{-2}$. 
This behavior is clearly related to the discontinuity in the electric field at $x=L$. 
In reality, the pair production front is 
not a sharp boundary, but rather a region of finite 
width over which the field drops continuously to zero. 
The width of this region
is determined by the distance over which the electric 
field can induce a substantial
charge separation in the newly created pairs. Assuming these are born 
with Lorentz factor $\gamma_{\rm pair}\approx \rho/L$, where $\rho$ is the 
radius of curvature of the 
trajectory imposed by the strong magnetic field
(see section~\ref{pulsars}), 
the width of the front can be estimated to be 
roughly $\Delta x\approx\gamma_{\rm pair}\accel\approx\accel\rho/L\gg\accel$. 
Thus, the emission produced at 
small beaming angles $\theta\sim\accel/L$ is unlikely to differ qualitatively
from emission at larger ones, and the entire pattern can be 
approximated as
a hollow cone, with a cut-off frequency that increases towards the inner
edge
\eqb
\label{Hyperestimate}
\frac{\diff U}{\diff\omega\diff\Omega}
&=&\left\lbrace
\begin{array}{ll}
\frac{32q^2}{9\pi^3c}
\left[\frac{\accel\omega}{c}K_1\left(\sin\theta\frac{\accel\omega}{c}\right)
\right]^2
&\textrm{for\ }\accel/L<\sin\theta<1\\
0 &\textrm{otherwise}
\end{array}
\right.
\eqe

Here, we have multiplied the spectrum (\ref{HyperAllmu}) by normalization
constant $32/9\pi$ such that the total
energy radiated matches that given by Larmor's formula (\ref{larmor}): 
\eqb
U=\frac{4}{3}\frac{q^2}{\accel}\sqrt{\gamma_{\rm max}^2-1}
\enspace.
\eqe

\begin{figure}
\includegraphics[width=0.45\textwidth]{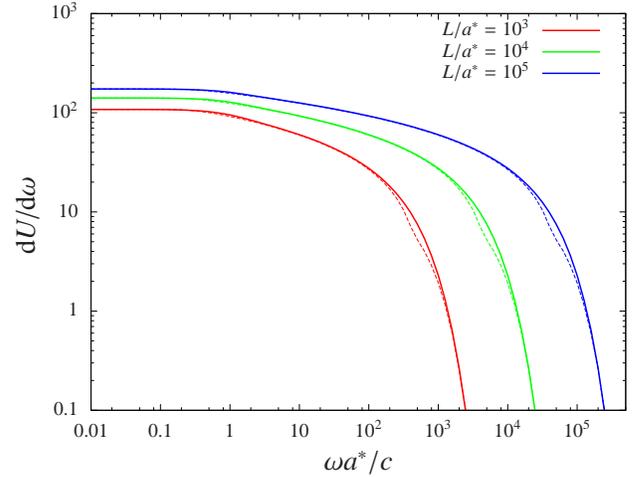}
\caption{
Comparison of numerical integration of eq. (\ref{Hyperestimate}) 
over solid angle (solid lines) with 
the approximate equation (\ref{approxhyper}) (dashed lines). A difference
of a few percent is found for $\omega\sim Lc/\accel^2$. 
The spectrum is again in units of $q^2/\pi^2c$.
A color version of this figure is available in the online journal.\\
 \label{HyperApproxFig}}
\end{figure}

If the details of the emission on angular scales $\accel/L$ are 
unimportant, a \lq\lq $\delta$-function\rq\rq\ approximation can be 
used. While the integration over solid angle cannot be found, a simple function
that approximates the angular integrated spectrum is 
\eqb
\label{approxhyper}
\frac{\diff U}{\diff \omega}=\frac{128q^2}{9\pi^2 c}{\rm Max}\left[
\ln\left(\frac{2L}{\accel}\tanh\frac{c}{3\omega\accel}\right),
\exp\left(-\frac{2\omega\accel^2}{Lc}\right)\right]\enspace.
\eqe
The accuracy of this approximation can be seen in Figure \ref{HyperApproxFig}
where the function is plotted together with the numerically integrated value
of (\ref{Hyperestimate}) for a range of $L/\accel$.

The maximum photon
energy produced can be reasonably approximated as
\eqb
\frac{\hbar\omega_{\rm max}}{m c^2} &=& L\Ehat/\accel
\nonumber\\
&=& L\Ehat^2/\lambdaC\enspace.
\label{hyperbolicmax}
\eqe
As we discuss in section \ref{pulsars}, this implies that 
each electron that traverses the gap from $x=0$ to $x=L$ produces
on average $\alpha_{\rm f}$ photons of frequency $\omega\sim\omega_{\rm max}$,
where $\alpha_{\rm f}$ is the fine-structure constant. 

\section{Spectra from motion in electrostatic waves}
\label{special}

\begin{figure}
\includegraphics[width=0.45\textwidth]{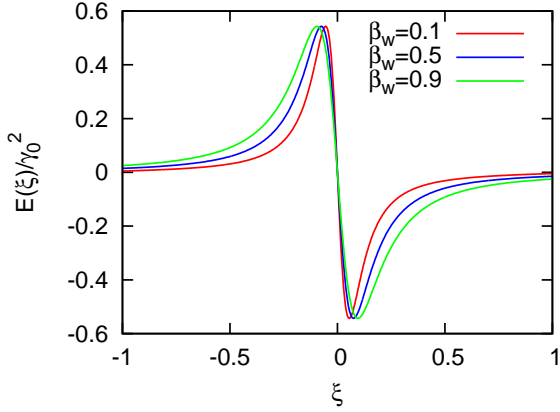}
\caption{
Electrostatic waves that yield an electric field at the particle position given
by  (\ref{ewave}), with $\gamma_0=10$, as a function of 
phase $\xi=\Omega_0(\beta_{\rm w}t+x/c)$ 
for three values of $\beta_{\rm w}$ (and assuming $q>0$). 
A color version of this figure is available in the online journal.\\
 \label{structures}}
\end{figure}

In this section we consider two special cases for which approximate 
or analytic expressions
can be found. The first is an isolated structure containing a reversal of the 
the electric field. To facilitate the analysis, we choose a particle orbit
of the form
\eqb
x(t)&=&\frac{c\beta_0}{\Omega_0}\textrm{tanh}
\left(\Omega_0 t\right)
\eqe
with $\left|\beta_0\right|<1$, i.e., the particle starts and 
finishes its orbit at rest, while its position suffers a displacement of
$2c\beta_0/\Omega_0$. The maximum particle speed is $c\beta_0$. The electric field
at the position of the particle is 
\eqb
E[x(t),t]&=&\frac{mc\Omega_0}{q}\left\lbrace\frac{-2\beta_0\textrm{tanh}(\Omega_0 t)\,\textrm{sech}^2(\Omega_0 t)}
{\left[1-\beta_0^2\textrm{sech}^4(\Omega_0 t)\right]^{3/2}}\right\rbrace\enspace.
\eqe
Of the many functions $E(x,t)$ that can provide
such a trajectory, those representing an isolated structure that is either 
static or moving at constant, subluminal velocity (which is nowhere equal 
to the 
particle speed) are perhaps the simplest that are physically
realistic. In the static case, the electric field is simply
\eqb
\label{efieldposition}
E(x)&=&
\left\lbrace
\begin{array}{ll}
\frac{mc\Omega_0}{q}
\frac{2\beta_0 \hat{x}\left(\hat{x}^2-1\right)}
{\left[1-\beta_0^2\left(1-\hat{x}^2\right)^2\right]^{3/2}}
&\textrm{for\ }\left|\hat{x}\right|\le1\\
0&\textrm{otherwise}
\end{array}
\right.
\label{ewave}
\eqe
where $\hat{x}=\Omega_0 x/c\beta_0$, but $E(x,t)$ 
must be constructed using a simple algorithm in the 
moving case. Examples of these structures are shown in Fig.~\ref{structures}.
If they accelerate particles to relativistic velocity, i.e., 
$\gamma_0=1/\sqrt{1-\beta_0^2}\gg1$, they
have the property that the maxima of the electric field are separated
in space by roughly $\Delta x\approx c/\gamma_0\Omega_0$ and have magnitude
$E_{\rm max}\approx \gamma_0^2 mc\Omega_0/q$.
Note that these are not self-consistent structures; they have been 
chosen to illustrate the properties of linear acceleration radiation.

From (\ref{schwinger2}), allowing the limits $t_{1,2}\rightarrow\pm\infty$,
one finds
\eqb
\label{dudodO}
\frac{\diff U}{\diff\Omega\diff\omega}&=&
\frac{q^2\omega^2\tan^2\theta}{4\pi^2c\Omega_0^2}
\left|\int_{-\infty}^{\infty}\diff t\exp\left[\frac{\omega}{\Omega_0}f(t)\right]
\right|^2\enspace,
\\
f(t)&=&it-i\cos\theta\,\beta_0\textrm{tanh}t\enspace.
\eqe
This integral can be approximated using the method of steepest descents.
As a function of complex argument $z=x+iy$, $f(z)$ has a suitable 
saddle-point at $x=0$, $y=\textrm{arccos}\sqrt{\beta_0\cos\theta}$ ($0<y<\pi/2$). The result is  
best expressed in terms of the angle $\bar{\theta}$ between the direction of 
propagation of the photon and the $x$-axis, as measured in a frame moving
along this axis at speed $\beta_0$, in which case
\eqb
\cos\bar{\theta}&=&\frac{\cos\theta - \beta_0}{1-\beta_0\cos\theta}
\eqe
and one arrives at the expression:
\eqb
\frac{\diff U}{\diff\Omega\diff\omega}&=&
\frac{3q^2\gamma_0^2}{8\pi c}\sin^2\bar{\theta}\frac{\omega}{\omega_c}
\exp\left[-\frac{\omega}{\omega_c}\right]\enspace.
\eqe

Thus, the spectrum rises linearly with frequency up to a cut-off
at 
\eqb
\omega_{\rm c}&=&3(1+\cos\bar{\theta})^{3/2}\gamma_0^3\Omega_0/4\enspace.
\eqe
Apart from the low frequency spectrum, which in any case is not reliably
given by the method of steepest descents, this spectrum 
is roughly the same as that emitted by a particle
experiencing uniform acceleration
in a field $E\approx E_{\rm max}=\gamma_0^2mc\Omega_0/q$ over a distance 
$L=\Delta x=c/\left(\gamma_0\Omega_0\right)$, given in (\ref{hyperbolicmax}).

\begin{figure}
\includegraphics[width=0.45\textwidth]{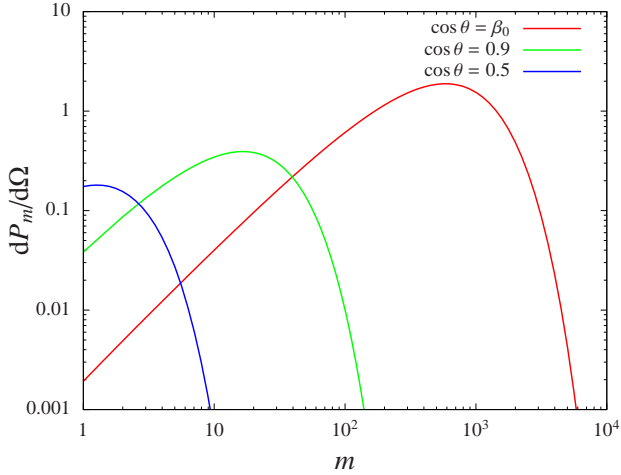}
\caption{
Envelope function for the power radiated in the $m$-th harmonic for different angles
in units of $q^2\Omega_0^2/2\pi c$ for an electron with maximum Lorentz factor $\gamma_0=10$,
c.f. Equation (\ref{Pmomega}).
A color version of this figure is available in the online journal.\\
 \label{spectrumJ}}
\end{figure}

The second case we consider is motion in a periodic wave. To find an 
analytic solution, we choose a particularly simple particle orbit
\eqb
x(t)=\frac{c\beta_0}{\Omega_0}\cos(\Omega_0 t)\enspace.
\eqe
I.e., the particle undergoes simple harmonic motion in one dimension.
The spectrum produced by a particle with such a trajectory has previously
been calculated by \citet{schott}.
The electric field at the position of the particle in this case is 
\eqb
\label{ewave2}
E[x(t),t]&=&\frac{mc\Omega_0}{q}\left\lbrace\frac{-\beta_0\cos(\Omega_0 t)}
{\left[1-\beta_0^2\sin^2(\Omega_0 t)\right]^{3/2}}\right\rbrace
\eqe
which, as in the previous case, has maxima of the electric field separated
in space by $\Delta x\approx c/\gamma_0\Omega_0$ with magnitude
$E_{\rm max}\approx \gamma_0^2 mc\Omega_0/q$. Thus, if the particle becomes highly relativistic, 
the field is large, but is confined to a small spatial region.
Since the motion of the particle is periodic, the spectrum consists of
harmonics of the oscillation frequency $\Omega_0$. 
The resulting power, averaged over one oscillation period, is
\eqb
\frac{\diff P}{\diff\Omega\diff\omega}&=&
\sum_{m=1}^{\infty}\delta(\omega-m\Omega_0)
\frac{\diff P_m}{\diff\Omega}
\enspace,
\label{Pmomega1}
\eqe
where 
\eqb
\frac{\diff P_m}{\diff\Omega}&=&
\frac{q^2\Omega_0^2\tan^2\theta}{2\pi c}
m^2J_m^2(\beta_0 m\cos\theta)\enspace.
\label{Pmomega}
\eqe
which clearly demonstrates the discrete nature of the spectrum.
An alternative derivation to that of \citet{schott} is given in the appendix.
Equation (\ref{Pmomega}) represents the average power radiated, in the $m$-th harmonic,
into a unit solid angle at an angle $\theta$ to the oscillation axis. 
For $m<m_c$ (see below) the power radiated in each harmonic at a fixed angle is
proportional to $m^{4/3}$. Fig.~\ref{spectrumJ} shows a plot of this function 
for different emission angles.

When $\omega$ is large, the distance $\Omega_0$ between neighboring harmonics becomes 
comparatively small, and the spectrum can be considered effectively as a continuum.
The transformation to a continuous frequency spectrum can be made by formally 
replacing the summation in equation (\ref{Pmomega1}) by an integral
$$\sum_{m=1}^{m=\infty}\longrightarrow\int_0^\infty\diff m$$
Taking the asymptotic value of the resulting expression, we find, 
in terms of the angle $\bar{\theta}$ measured
in a frame moving at speed $\beta_0$ 
\eqb
\left\langle
\frac{\diff U\left(\mbox{$\omega$}\right)}
{\diff\omega\diff\Omega}\right\rangle_{\rm period}\cong
\frac{3q^2\gamma_0^2}{16\pi c}\sin^2\bar{\theta}\frac{\omega}{\omega_{\rm c}}
\exp\left[-\frac{\omega}{\omega_{\rm c}}\right]
\eqe
where we have multiplied by a factor
$2\pi/\Omega_0$ to give the average energy emitted in each period,
The critical frequency in this case is
\eqb
\omega_{\rm c}&=&\Omega_0m_{\rm c}=\frac{3}{2^{5/2}}(1+\cos\bar{\theta})^{3/2}\gamma_0^3\Omega_0.
\eqe
This result is very similar to the approximate 
expression for an isolated structure, provided the strength of the 
electric field and its spatial extent coincide.

Approximate expressions for the emission
produced by a relativistic oscillator have been found by
\citet{wagoner69} and \citet{melroseluo09}. These 
share several properties with
the simpler cases presented here. 
To compare with the emission from a hyperbolic trajectory, note that 
the electric field varies only linearly over the wave period, so that 
its effective spatial extent is  $\Delta x \approx c/\Omega_0$.
The effective field strength, on the other hand 
is $E_{\rm max}\approx \gamma_{\rm max}mc\Omega_0/q$.
According to (\ref{hyperbolicmax}) a cut-off is expected at 
$\omega_{\rm crit} = \gamma_{\rm max}^2\Omega_0$, as found by the above authors.
The low frequency spectrum, however, is more 
sensitive to the wave form, with 
the oscillator having $\diff U/\diff\omega\diff\Omega\propto \omega$, whereas
the low frequency spectrum radiated by the hyperbolic orbit is flat.


\section{Application to pulsars}
\label{pulsars}

Pulsars are thought to lose their rotational energy in a relativistic
wind that contains a large number of electron-positron pairs
\citep[for a review, see][]{kirkpetrilyubarsky09}. 
These pairs presumably materialize in an electromagnetic cascade
initiated by electrons that are accelerated in the strong 
inductive electric fields surrounding the star. 
Recent observations indicate that 
gamma-ray pulses are unlikely to be produced close to the star 
\citep{abdo_catalog09,abdo_millisecond09}. However,
this is not necessarily the region where the pairs materialize,  
the location of which is still unknown. Detailed models of the electromagnetic 
cascades have been constructed 
\citep[reviewed in][]{harding09,cheng09}, but
these face difficulties in accounting for the observations
\citep{abdo_millisecond09,dejager07,timokhin09}. 

The radiation processes thought to be important in pulsar gap models
are curvature radiation and inverse Compton scattering (both resonant 
and non-resonant). Linear acceleration emission is closely related to 
curvature radiation, since they both depend on the 
configuration of the large-scale electromagnetic fields anchored in the star.
Inverse Compton 
scattering, on the other hand, depends 
not only on the local magnetic field strength, but also
on the target photon population.
Consequently, to assess the possible role of linear acceleration emission, 
we compare its spectrum and luminosity to the standard expressions describing
curvature radiation. 

The acceleration region or \lq\lq gap\rq\rq\ can be 
characterized by its height $L$, and the 
potential drop $\gamma_{\rm max}mc^2/e$ across it. 
Assuming the electric field to be uniform, it can
be expressed in units of the Schwinger field as
\eqb
\Ehat&=&\lambdaC\gamma_{\rm max}/L\enspace.
\eqe
Denoting by $\rho$ the radius of curvature of 
the magnetic field lines in the gap, the 
characteristic or cut-off frequencies of 
linear acceleration emission, $\omega_{\rm linac}=mc^2\omegalin/\hbar$,
and curvature emission,
$\omega_{\rm curv}=mc^2\omegacurv/\hbar$, are
\eqb
\omegalin&=&\gamma_{\rm max}\Ehat\,=\,\gamma_{\rm max}^2\lambdaC/L\enspace,
\\
\omegacurv&=&\gamma_{\rm max}^3\lambdaC/\rho\,=\,\omegalin
\left(\gamma_{\rm max}L/\rho\right)\enspace.
\eqe
The energy radiated 
as linear acceleration emission, $U_{\rm linac}=\hat{U}_{\rm linac}mc^2$ 
and as curvature 
radiation $U_{\rm curv}=\hat{U}_{\rm curv}mc^2$
by a single electron that traverses the gap is 
\eqb
\label{Uhat1}
\hat{U}_{\rm linac}&=&4\alpha_{\rm f}\gamma_{\rm max}\Ehat/3\,=\,
4\alpha_{\rm f}\gamma_{\rm max}^2\lambdaC/3L\enspace,
\\
\label{Uhat2}
\hat{U}_{\rm curv}&=&2\alpha_{\rm f}\gamma_{\rm max}^4\lambdaC L/3\rho^2
\,=\,\mbox{$\frac{1}{2}$} \hat{U}_{\rm linac}\left(\gamma_{\rm max}L/\rho\right)^2\enspace.
\eqe

These expressions show that linear acceleration emission and
curvature radiation are complementary rather than competing processes.
Each of them is an approximate evaluation of 
the radiation emitted by a charge accelerated along a
curved trajectory, computed according to classical electrodynamics.
In the case of linear acceleration emission, the approximation of large
radius of curvature is used. In the case of curvature emission, the particle's
acceleration parallel to its velocity is neglected. 
If the system is such that $L>\rho/\gamma_{\rm max}$, i.e., if 
the formation length for curvature photons is smaller than the acceleration
region, then the radiation is well described by a local emissivity, 
and this can be evaluated from the standard curvature formulas, 
because the change in
Lorentz factor is small within the photon formation length. 
On the other hand, if 
$L<\rho/\gamma_{\rm max}$, the photon formation length is roughly equal to 
the size of the system and it is impossible to define a local emissivity.
In this case, it is inconsistent to use the expressions derived for
curvature radiation, since they apply only when the formation length 
is limited by the curvature of the trajectory.  
When the formation length is instead limited 
by the size of the acceleration region, the expressions derived for 
linear acceleration emission are applicable.

Note that these arguments do not apply to particles that are accelerated
externally and are injected into 
the gap with $\gamma>\gamma_{\rm max}$. In this case,
the effects of the longitudinal electric field are small and can be 
treated using a linear theory \citep{melrose78}. Strictly speaking, 
the arguments also do not apply if the particle is accelerated after leaving 
the gap --- for example, if it continues to propagate along a 
curved field line without linear acceleration. 
However, in this case the local emissivity for curvature radiation can be used 
to describe the additional radiation.

Note that the average number of photons
emitted by a particle as it traverses a photon formation length is 
$\alpha_{\rm f}$. This is true both for 
linear acceleration emission and curvature 
(and synchrotron) radiation. 
For a classical picture to be valid, the energy of the radiated photon 
must be small compared to the energy of the radiating particle.
In the case of linear
acceleration emission this implies $\Ehat\ll1$. 
Therefore, since the energy radiated is 
$\sim\alpha_{\rm f}\Ehat\gamma_{\rm max}mc^2\ll \gamma_{\rm max}mc^2$, 
classical linear acceleration emission cannot lead to a saturation of the 
acceleration process,
in the sense of limiting $\gamma_{\rm max}$. This is consistent with the fact
that the radiation reaction force vanishes at all points on the trajectory 
except the end points. Classical curvature 
radiation, on the other hand, does saturate the acceleration, if the 
region exceeds a size $L_{\rm max}$ given by
\eqb
L_{\rm max}&=&\left(\rho/\gamma_{\rm max}\right)
\left(\alpha_{\rm f}\Ehat\right)^{-1/2}\enspace.
\eqe

Gaps in pulsar magnetospheres can be divided into two types: vacuum gaps
\citep{sturrock71,rudermansutherland75} and
gaps produced by space-charge limited flow \citep{aronsscharlemann79}. 
The electric field is produced by the rotation of the strong surface 
magnetic field $B_{\rm surface}=\Bhat B_{\rm crit}$ with 
$B_{\rm crit}=m^2c^3/e\hbar=4.414\times10^{13}\,$G. 
In vacuum gaps, it may be as high as $B_{\rm surface}$ 
times the rotation speed/$c$, and it is convenient to express it 
in terms of this value
by introducing 
the parameter $\evac\sim1$ according to
\eqb
\Ehat&=&\evac\Bhat \rstar
\nonumber\\
&\approx&\evac\,
5\times10^{-6}\left(\Pdot/P\right)^{1/2},\ \textrm{(vacuum)}
\label{Evac}
\eqe
where $\rstar=R_*/\rlight$, 
is
the ratio of the stellar radius $R_*$ to the light-cylinder radius
$\rlight$,
$P$ is the pulsar period measured in seconds, 
and we have used the standard estimate for the surface magnetic field from 
the rate of loss of rotational energy  
\eqb
\Bhat&\approx&2.3\times10^{-2}\left(P\Pdot\right)^{1/2}
\eqe
where $\Pdot$ is the rate of change of pulsar period times
$10^{15}$.

The presence of space-charge limits the electric field 
to much smaller values. When 
the effects of frame-dragging are taken into account 
\citep{muslimovtsygan92,muslimovharding97} one finds
\eqb
\Ehat&\approx&\espace\,0.1\Bhat\rstar^2
\nonumber\\
&\approx&\espace\,10^{-10}\Pdot^{1/2}P^{-3/2},\ \textrm{(space-charge)}
\label{Espace}
\eqe
where, in analogy with (\ref{Evac}), we have introduced the parameter
$\espace\sim1$.

The ratio $L\gamma_{\rm max}/\rho$
of the system size to the formation length for curvature photons determines which radiation mechanism
is appropriate in a particular gap.
This condition can be formulated in terms of a critical Lorentz factor:
\eqb
\gamma_{\rm crit}&=&\sqrt{\Ehat \rho/\lambdaC}
\nonumber\\
&=&
\left\lbrace
\begin{array}{ll}
2.4\times10^7 \left(\evac\rhostar\right)^{1/2}
P^{1/4}\Pdot^{1/4}&\textrm{(vacuum)}
\\
1.1\times10^5 \left(\espace\rhostar\right)^{1/2}
P^{-1/4}\Pdot^{1/4}&\textrm{(space-charge)}
\end{array}
\right.
\label{gcrit}
\eqe
where we have written the radius of curvature of the field lines in units
of the light cylinder radius, $\rho=\rhostar\rlight$, and have inserted the 
expressions (\ref{Evac}) and (\ref{Espace}) for the accelerating fields.
If $\gamma_{\rm max}>\gamma_{\rm crit}$  
($\gamma_{\rm max}<\gamma_{\rm crit}$), 
curvature radiation (linear acceleration emission) is the appropriate
description. Alternatively, the curvature photon formation length $L_{\rm curv}$ 
can be evaluated:
\eqb
L_{\rm curv}&=&\rho/\gamma_{\rm crit}
\nonumber\\
&=&
\left\lbrace
\begin{array}{ll}
200\left(\evac\rhostar\right)^{1/2}
P^{3/4}\Pdot^{-1/4}\,\textrm{cm}&\textrm{(vacuum)}
\\
4\times10^4 \left(\espace\rhostar\right)^{1/2}
P^{5/4}\Pdot^{-1/4}\,\textrm{cm}&\textrm{(space-charge)}
\end{array}
\right.
\label{hcrit}
\eqe

According to (\ref{hcrit}), curvature radiation is an appropriate description
of the radiation from a vacuum gap
(space-charge flow limited gap) provided its height exceeds 
roughly $200\,\textrm{cm}$ ($4\times10^4\,\textrm{cm}$).
Note, however, that these conditions are relaxed (in the sense that the 
critical gap size is reduced) 
if the radius of curvature of the field lines is small
compared to the light-cylinder radius, and also 
if the accelerating field fails to 
reach the full value given by (\ref{Evac}) or (\ref{Espace}) ---
both of which were assumed in the original
\citet{rudermansutherland75} model.

In current static models, the assumed gap heights greatly exceed
$L_{\rm curv}$, so that linear acceleration emission does not play a
role.
Recently, however, time-dependent gap models have been proposed
\citep{levinsonetal05,timokhin09}. These models assume 
accelerating fields that are of the order of, or larger than the 
vacuum field given in (\ref{Evac}). In addition, 
time-dependent screening leads to structure on a 
length scale determined essentially by the 
electron inertial length $\ell_{\rm e}=c/\omega_{\rm pe}$, where
$\omega_{\rm pe}$ is the local plasma frequency. Estimating the 
electron/positron charge 
density as roughly equal to the charge-density required to screen 
the electric field (the \lq\lq Goldreich-Julian\rq\rq\ density) implies
\eqb
\omega_{\rm pe}&=&1.5\times10^{10}\,P^{-1/4}\dot{P}^{1/4}\,\textrm{s}^{-1}
\eqe
so that
\eqb
\frac{\ell_{\rm e}}{L_{\rm curv}}
&=&
\left\lbrace
\begin{array}{ll}
10^{-2}\left(\evac\rhostar\right)^{-1/2}
P^{-1/2}&\textrm{(vacuum)}
\\
5\times10^{-5}\left(\espace\rhostar\right)^{-1/2}
P^{-1}&\textrm{(space-charge)}
\end{array}
\right.
\label{hcrit2}
\eqe
Therefore, linear acceleration emission is 
an essential ingredient in these models, because the 
formation length for curvature radiation photons
substantially exceeds the anticipated length-scale of the
structure in the accelerating electric field.

\section{Conclusions}
\label{conclusions}

The radiation produced by a particle whose acceleration is 
parallel to its velocity, linear acceleration emission,
has been investigated for a number of special cases. 
Particle motion of this type is expected to occur in the intense magnetic
fields close to the surface of a pulsar.
The emission produced by these particles is conceptually difficult
due to the macroscopic size of the photon formation length. 
For the particular case of hyperbolic motion, corresponding to
acceleration in a uniform electric field, the formation length is
comparable to the length of the entire accelerating region. 
Linear acceleration emission is similar to both synchrotron and curvature radiation
in the sense that a photon is emitted with a probability $\alpha_{\rm f}$
when a particle traverses a formation length.
The essential properties of the radiation 
produced by a particle on a hyperbolic trajectory can be understood in terms of
two key parameters: the length $L$ of the hyperbolic part of the trajectory,
and the length $\accel$ over which the particle increases its energy
by $mc^2$. Some generic features of the
emission mechanism in different electric field configurations can also be 
described using these quantities. In particular, provided  $L$ and $\accel$ are 
interpreted appropriately (see section \ref{special} for some examples) 
the critical frequency can be expressed as $\omega_{\rm crit}\approx Lc/\accel^2$.

Current models of pair production in pulsar magnetospheres typically
assume that high energy photons that
induce the secondary pair cascade, are produced via either
curvature radiation or inverse-Compton scattering.  An alternative to
these processes is linear acceleration emission, as investigated by
\citet{melroseluo09}. We have demonstrated that this process is in
fact complementary to curvature radiation, the transition between the
two regimes depending only on the ratio of the system size to the
formation length of curvature photons.  For existing static polar gap
models, linear acceleration is unlikely to play an important role,
although this relies on the environmental
parameters being close to the canonical values. The limits for both vacuum gap
models and space-charge limited models are presented in section
\ref{pulsars}.

Recent numerical studies provide evidence that the above stationary models are
unstable to perturbations \cite[e.g.][]{levinsonetal05, timokhin09}. Several 
analytic models have been developed in which 
particles are accelerated in large amplitude electrostatic waves that propagate
parallel to the magnetic field \cite[see e.g.][and references therein]{luomelrose08}.
Since the characteristic length scale of the electric fields in these models is on
the order of the electron inertial length, linear acceleration emission is likely
to be a more appropriate description of the resulting radiation.

\acknowledgements
We thank Don Melrose, Qinghuan Luo and Mohammed Rafat for stimulating discussions.
B.R. gratefully acknowledges support from the Alexander von Humboldt foundation.

\appendix

\label{AppendixA}

Here we provide an alternative derivation to that of \citet{schott} 
for the spectrum radiated by a particle
undergoing simple harmonic motion.
Taking advantage of the periodic nature of the orbit, Eq. (\ref{emissivity})
can be rewritten in terms of an infinite sum
\eqb
\frac{\diff U}{\diff\Omega\diff\omega}&=&
\frac{q^2\tan^2\theta}{4\pi^2c}
\left|\sum_{k=-\infty}^{k=\infty}\frac{\omega}{\Omega_0}{\rm e}^{2\pi i k\frac{\omega}{\Omega_0}}
\int_{-\pi}^{\pi}\diff\phi{\rm e}^{i\frac{\omega}{\Omega_0}\left(\phi-\mu\beta_0\cos\phi\right)}
\right|^2
\eqe
where $\phi=\Omega_0 t+2\pi k$. Using Poisson's sum formula
\eqb
\sum_{k=-\infty}^{k=\infty}{\rm e}^{2\pi i k \frac{\omega}{\Omega_0}} = 
\sum_{m=-\infty}^{m=\infty}\delta\left(\frac{\omega}{\Omega_0}-m\right) 
\eqe
and noting that $\omega$ is positive, we find 
\eqb
\frac{\diff U}{\diff\Omega\diff\omega}&=&
\frac{q^2\Omega_0^2\tan^2\theta}{4\pi^2c}
\left|\sum_{m=1}^{m=\infty}\delta(\omega-m\Omega_0)m
\int_{-\pi}^{\pi}\diff\phi{\rm e}^{im\left(\phi-\mu\beta_0\cos\phi\right)}
\right|^2
\eqe
The integral in this equation can be expressed in terms the Bessel functions
\eqb
\frac{\diff U}{\diff\Omega\diff\omega}&=&
\frac{q^2\Omega_0^2\tan^2\theta}{c}
\left|\sum_{m=1}^{m=\infty}\delta(\omega-m\Omega_0)m
i^mJ_m(\mu\beta_0 m)
\right|^2
\eqe
where we have exploited the integral form of the Bessel function
\eqb
i^mJ_m(t)=\int_0^{2\pi}\frac{\diff\phi}{2\pi}{\rm e}^{i(t\cos\phi-m\phi)}
\eqe
Finally using the identity
\eqb
\delta^2(x)=\lim_{T\rightarrow\infty}\frac{T}{2\pi}\delta(x)
\eqe
the power radiated in the $m$-th harmonic is
\eqb
\frac{\diff P}{\diff\Omega\diff\omega}&=&\lim_{T\rightarrow\infty}
\frac{1}{T}\frac{\diff U}{\diff\Omega\diff\omega}=
\sum_{m=1}^{m=\infty}\delta(\omega-m\Omega_0)
\frac{\diff P_m}{\diff\Omega}
\eqe
with
\eqb
\frac{\diff P_m}{\diff\Omega}&=&
\frac{q^2\Omega_0^2\tan^2\theta}{2\pi c}
m^2J_m^2(\mu\beta_0 m).
\eqe
The validity of this result can be checked by comparing against
Larmor's formula Eq. (\ref{larmor}). This can be done straightforwardly using the identity
\eqb
\sum_{m=1}^{\infty}m^2 J_m(mx)=\frac{x^2(4+x^2)}{16(1-x^2)^{7/2}}
\eqe
and subsequently integrating over solid angle. Multiplying the resulting answer 
by the period of the waves $2\pi/\Omega_0$ gives the total energy radiated
per oscillation 
\eqb
U=\frac{\pi q^2 \Omega_0\beta_0^2}{6 c}\gamma_0^3
\left[1+\frac{3}{\gamma^2}\right]. 
\eqe
It is easily verified that this agrees exactly with Eq. (\ref{larmor}).

The asymptotic expression for Bessel functions of large order and argument
is \citep[e.g][p. 249]{Watson}
\eqb
J_m(x)\sim\frac{1}{\pi}\sqrt{\frac{2(m-x)}{3x}}K_{1/3}\left[\frac{2^{3/2}(m-x)^{3/2}}{3x^{1/2}}\right]
\eqe
with $x<m$. Thus, using the asymptotic expression for the modified Bessel functions with
large argument, the spectrum at high harmonics is
\eqb
\frac{\diff P_m}{\diff\Omega}&=&
\frac{q^2\Omega_0^2}{2\pi c}
\frac{\tan^2\theta}{\sqrt{2}\sqrt{\mu\beta_0}\sqrt{1-\mu\beta_0}}
m\exp\left[-\frac{2^{5/2}(1-\mu\beta_0)^{3/2}}{3\sqrt{\mu\beta_0}}m\right]
\eqe

\end{document}